\documentclass[10pt, twocolumn]{IEEEtran}
\IEEEoverridecommandlockouts
\usepackage{cite}
\usepackage{amsmath,amssymb,amsfonts}
\usepackage{algorithmic}
\usepackage{graphicx}
\usepackage{textcomp}
\def\BibTeX{{\rm B\kern-.05em{\sc i\kern-.025em b}\kern-.08em
    T\kern-.1667em\lower.7ex\hbox{E}\kern-.125emX}}
\usepackage{amsmath}
\usepackage{amsmath,amsthm}
\usepackage{cite}   
\usepackage{graphicx}  
\usepackage{amsmath, amsopn, psfrag, amsthm}   
\usepackage{amssymb}
\usepackage{color}
\usepackage{algorithm}
\usepackage{amsmath}
\usepackage{algorithmic}
\usepackage[utf8]{inputenc}
\usepackage[english]{babel}
\newtheorem{theorem}{Theorem}
\newtheorem{lemma}{Lemma}

\newtheorem{proposition}{Proposition}
\usepackage{psfrag}
\usepackage{epsfig}
\graphicspath{{./img/}}
\DeclareGraphicsExtensions{.pdf,.jpeg,.png, .eps}
\DeclareMathOperator*{\argmax}{arg\, max}

\usepackage{amssymb,amsmath,mathtools}

\def\bgama{{\boldsymbol{\Gamma}}}
\def\bPhi{{\boldsymbol{\Phi}}}
\def\bep{{\boldsymbol{\epsilon}}}

\def\bdel{{\boldsymbol{\Delta}}}
\usepackage{cite}   
\usepackage{graphicx}  
\usepackage{amsmath, amsopn}   
\usepackage{amssymb}
\usepackage{color}
\usepackage{psfrag}
\usepackage{epsfig}
\usepackage{stfloats}
\usepackage{lipsum}
\usepackage{multicol}
\graphicspath{{./img/}}
\DeclareGraphicsExtensions{.pdf,.jpeg,.png}
\begin{document}
\title{On the Energy Efficiency of Limited-Backhaul Cell-Free Massive MIMO}%
\vspace{-1.2cm}
\linespread{.89}
\vspace{-2cm}
\author{
\IEEEauthorblockN{Manijeh Bashar\IEEEauthorrefmark{1}\IEEEauthorrefmark{4}, Kanapathippillai Cumanan\IEEEauthorrefmark{1}, Alister G. Burr\IEEEauthorrefmark{1}, Hien Quoc Ngo\IEEEauthorrefmark{2}, Erik G. Larsson\IEEEauthorrefmark{3},and 
Pei Xiao\IEEEauthorrefmark{4}}

\IEEEauthorblockA{\IEEEauthorrefmark{1}University of York, UK, \IEEEauthorrefmark{2}Queen's University Belfast, UK, \IEEEauthorrefmark{3}Linköping University, Sweden,
\IEEEauthorblockA{\IEEEauthorrefmark{4}Home of the 5G Innovation Centre, Institute for Communication Systems, University of Surrey, UK.}}
Email:$\!${\{$\!$mb1465,kanapathippillai.cumanan,alister.burr$\!$\}$\!$@$\!$york.ac.uk},\\hien.ngo@$\!$qub.ac.uk,erik.g.larsson$\!$@$\!$liu.se,p.xiao@$\!$surrey.ac.uk
}
\markboth{IEEE ICC,~2019}
{Shell \MakeLowercase{\textit{et al.}}: Bare Demo of IEEEtran.cls for IEEE Communications Society Journals}
\vspace{-.2cm}
\maketitle
\begin{abstract}
We investigate the energy efficiency performance of cell-free Massive multiple-input multiple-output (MIMO), where the access points (APs) are connected to a central processing unit (CPU) via limited-capacity links. Thanks to the distributed maximum ratio combining (MRC) weighting at the APs, we propose that only the quantized version of the weighted signals are sent back to the CPU. Considering the effects of channel estimation errors and using the Bussgang theorem to model the quantization errors, an energy efficiency maximization problem is formulated with per-user power and backhaul capacity constraints as well as with throughput requirement constraints. To handle this non-convex optimization problem, we decompose the original problem into two sub-problems and exploit a successive convex approximation (SCA) to solve original energy efficiency maximization problem. Numerical results confirm the superiority of the proposed optimization scheme.
\end{abstract}
\vspace{-.2cm}
\section{Introduction}
 \let\thefootnote\relax\footnotetext{The work of K. Cumanan and A. G. Burr was supported by H2020- MSCA-RISE-2015 under grant number 690750. This work was also supported in part by the U.K. Engineering and Physical Sciences  Research  Council  under Grant EP/N020391/1. The work of H. Q. Ngo was supported by the UK Research and Innovation Future Leaders Fellowships under Grant MR/S017666/1.}
One of the main issues for cell-free Massive multiple-input multiple-output (MIMO) systems is the limited capacity of the links from the access points (APs) to the central processing unit (CPU). Following \cite{marzetta_free16,eehienfree,ouricc1,ourjournal1,alister_pimrc_18,our_icc19_noma}, we refer to these links as \textit{backhaul links}. We study the case when only the quantized version of the weighted signal is available at the CPU and the CPU employs maximum ratio combining (MRC) detection. In \cite{ouricc2, ourjournal2,ourvtc18,ourasilomar18,dick_tvt_19}, the authors show that exploiting optimal uniform quantization and wireless microwave links with capacity 100 Mbits/s, the performance of limited-backhaul cell-free Massive MIMO system closely approaches the performance of cell-free Massive MIMO with perfect backhaul links.
We consider the energy efficiency maximization problem, where to tackle the non-convexity of the optimization problem, we decouple the original problem into two sub-problems, namely, receiver filter coefficient design, and power allocation. We next show that the receiver filter coefficient design problem can be solved through a generalized eigenvalue problem \cite{bookematrix,
our_lett_18,
cuma_jointbf_twc10,
cuma_mmse_tvt13,cuma_rate_icc11,
mythesis,our_tgcn_accepted}. Unfortunately, the user power allocation problem is a non-convex problem, and a successive convex approximation (SCA) is used to convert the original power allocation problem into a geometric programming (GP) problem. This scheme introduces an efficient solution for the original problem \cite{otterson_ee_15,refrence52ofotterson_ee_15}.
The contributions of the paper are summarized as follows: 
\begin{itemize}
\item[\textbf{1.}] A closed-form expression for the energy efficiency is derived, where we exploit the Bussgang
decomposition to model the effect of quantization, and present the analytical solution to find the optimal step-size of the quantizer. An expression for uplink energy efficiency is derived based on channel statistics and taking into
account the effects of channel estimation errors, the effect of pilot sequences, and quantization error.

\item[\textbf{2.}] We decompose the non-convex original problem into two sub-problems and an iterative algorithm is developed to determine the optimal solution. An SCA is used to efficiently solve the power allocation problem. Numerical results demonstrate that the proposed scheme substantially outperforms the case with equal power allocation. 
\end{itemize}
\vspace{-.1cm}        
\section{SYSTEM MODEL}
\vspace{-.1cm}
We consider uplink transmission in a cell-free Massive MIMO system with $M$ APs and $K$ randomly distributed single-antenna users in a large area. Moreover, we assume each AP has $N$ antennas. The channel coefficients between the $k$th user and the $m$th AP, $\textbf{g}_{mk} \in \mathbb{C}^{N\times 1}$, is modeled as
$
\textbf{g}_{mk}=\sqrt{\beta_{mk}}\textbf{h}_{mk},
$
where $\beta_{mk}$ denotes the large-scale fading and the elemnts of $\textbf{h}_{mk}$ are independent and identically distributed (i.i.d.) $\mathcal{CN}(0,1)$ random variables, and represent the small-scale fading \cite{marzetta_free16}.
All pilot sequences transmitted by the $K$ users in the channel estimation phase are collected in a matrix $\bPhi \in \mathbb{C}^{\tau_p\times K}$, where $\tau_p$ is the length of the pilot sequence (in symbols) for each user and the $k$th column of $\bPhi$, $\pmb{\phi}_k$, represents the pilot sequence used for the $k${th} user. After performing a de-spreading operation, the minimum mean square error (MMSE) estimate of the channel coefficient between the $k$th user and the $m$th AP is given by \cite{marzetta_free16}
\vspace{-.11cm}
\begin{IEEEeqnarray}{rCl}
\vspace{-.29cm}
	\hat{\textbf{g}}_{mk}\!\!=\!\!c_{mk}\!\left(\!\!\!\sqrt{\tau_p p_p}\textbf{g}_{mk}\!\!+\!\!\sqrt{\tau_p p_p}\sum_{k^\prime\ne k}^{K}\textbf{g}_{mk^\prime}\pmb{\phi}_{k^\prime}^H\pmb{\phi}_{k}\!\!+\!\!\textbf{W}_{p,m}\pmb{\phi}_k\!\!\right)\!\!,~
	\label{ghat}
\end{IEEEeqnarray}
where $\textbf{W}_{p,m}$ denotes the noise sequence at the $m$th antenna  whose elements are independent and identically distributed (i.i.d) $\mathcal{CN}(0,1)$, $p_p$ represents the normalized signal-to-noise ratio (SNR) of each pilot symbol (which we define in Section VI), and $c_{mk}$ is given by
$
c _{mk}=\frac{\sqrt{\tau_p p_p}\beta_{mk}}{\tau_p p_p\sum_{k^\prime=1}^{K}\beta_{mk^\prime}|\pmb{\phi}_k^H{\pmb{\phi}}_{k^\prime}|^2+1}.
$  The investigation of cell-free Massive MIMO with realistic COST channel model \cite{ourglobecom_cost,tvt_cost_me,our_ew} will be considered in our future work. 
\vspace{-.2cm}
\subsection{Uplink Transmission}
\vspace{-.05cm}
The transmitted signal from the $k$th user is represented by
$
x_k= \sqrt{q_k}s_k,
$
where $s_k$ ($\mathbb{E}\{|s_{k}|^2\} = 1$) and $q_k$ denotes the transmitted symbol and the transmit power from the \textit{k}th user, respectively. 
The received signal at the $m$th AP is given by 
\vspace{-.22cm}
\begin{equation}
\textbf{y}_m= \sqrt{\rho}\sum_{k=1}^{K}\textbf{g}_{mk}\sqrt{q_k}s_k+\textbf{n}_m, 
\label{ym}
\end{equation}
where each element of $\textbf{n}_m \in \mathbb{C}^{N\times 1}$, $n_{n,m}\sim \mathcal{CN}(0,1)$ is the noise at the $m$th AP, and $\rho$ is the normalized uplink SNR.
\vspace{-.22cm}
\subsection{Optimal Uniform Quantization Model}
\label{sec_ee_bussa}
\vspace{-.2cm}
The Bussgang theorem \cite{Zillmann} is exploited, where a nonlinear output of a quantizer can be introduced by a linear function plus uncorrelated distortion as
$
\mathcal{Q} (z) = az+n_d, ~\forall k, 
$
where $a$ is a constant, $n_d$ refers to the distortion noise, $z$ is the input of the quantizer\cite{Zillmann,ourvtc18,ourjournal2}. The term $a$ is given by 
$
a=\frac{\mathbb{E}\left\{zh(z)\right\}}{\mathbb{E}\{z^2\}}=\frac{1}{p_z}\int_{\mathcal{Z}}zh(z)f_z(z)d~z ,
$
where $p_z=\mathbb{E}\{|z|^2\}=\mathbb{E}\{z^2\}$ denotes the power of $z$ and we drop absolute value as $z$ is a real number, and $f_z(z)$ represents the probability distribution function of $z$. We define the second parameter 
$
b=\frac{\mathbb{E}\left\{h^2(z)\right\}}{\mathbb{E}\{z^2\}}=\frac{1}{p_z}\int_{\mathcal{Z}}h^2(z)f_z(z)dz
$ \cite{Zillmann,ourvtc18,ourjournal2}. We aim to maximize the signal-to-distortion noise ratio (SDNR), which is defined as follows:
$
\text{SDNR}=\frac{\mathbb{E}\left\{(az)^2\right\}}{\mathbb{E}\{n_d^2\}}=\frac{a^2}{b-a^2},
$
where $\mathbb{E}\left\{az^2\right\}=a^2p_z$, and $\mathbb{E}\{n_d^2\}=p_{n_d}=(b-a^2)p_z$.
In practice, we divide the input by its standard deviation, and multiply the output by the same factor. By introducing a new variable $\tilde{z}=\frac{z}{\sqrt{p_z}}$, we have 
\vspace{-.22cm}
\begin{equation}
\mathcal{Q}(z)=\sqrt{p_z}\mathcal{Q}(\tilde{z})=\tilde{a}\sqrt{p_z}\tilde{z}+\sqrt{p_z}\tilde{n}_d=\tilde{a}z+\sqrt{p_z}\tilde{n}_d.
\label{sigmamultiplyq}
\end{equation}
The optimal step-size of the quantizer, $\Delta_{\text{opt}}$, can be obtained by solving the following maximization problem:
$
\Delta_\text{opt}=\argmax_{\Delta}~{\text{SDNR}}.
$
In \cite{ourvtc18,ourjournal2}, by deriving closed-form expressions for $a$ and $b$, we numerically solve this maximization problem, and the resulting distortion power are summarized in Table \ref{tablezillmann}, where $\alpha$ refers to the number of quantization bits.
\vspace{-.11cm}
\subsection{The Signal Received at the CPU}
\vspace{-.12cm}
At each AP, MRC weighting is performed. Using Bussgang's theorem \cite{Zillmann}, a nonlinear output can be represented as a linear function as follows:
$
\mathcal{Q} \left(\mathcal{R}\left(\hat{\textbf{g}}_{mk}^{H}\textbf{y}_m\right)\right)
 = \tilde{a}\mathcal{R}\left(\hat{\textbf{g}}_{mk}^{H}\textbf{y}_m\right)+\sigma_{\mathcal{R}\left(\hat{\textbf{g}}_{mk}^{H}\textbf{y}_m\right)}\tilde{n}_{d,mk}, ~\forall k, 
$
where $\sigma_{\mathcal{R}\left(\hat{\textbf{g}}_{mk}^{H}\textbf{y}_m\right)}$ is the standard deviation of the $\mathcal{R}\left(\hat{\textbf{g}}_{mk}^{H}\textbf{y}_m\right)$, where $\mathcal{R}$ represents the real part of a complex number.
Note that as mentioned in the previous subsection, we use the scheme in \cite{ourvtc18} to exploit Bussgang decomposition. Here, given the fact that the input of quantizer, i.e., $\hat{\textbf{g}}_{mk}^{H}\textbf{y}_m$, is the summation of many terms, it can be approximated as a Gaussian random variable. This enables us to exploit the values given in Table \ref{tabledelta}, which are obtained for Gaussian input.
 Note that
$
\sigma_{\mathcal{R}\left(
\hat{\textbf{g}}_{mk}^{H}\textbf{y}_m\right)}^2=\sigma_{\mathcal{I}\left(\hat{\textbf{g}}_{mk}^{H}\textbf{y}_m\right)}^2=\frac{1}{2} \sigma_{\hat{\textbf{g}}_{mk}^{H}\textbf{y}_m}^2.
$To improve the performance, the signal is further multiplied by the receiver filter coefficients $u_{mk}, \forall m,k$ at the CPU. The received signal at the CPU can be written as
\begin{IEEEeqnarray}{rCl}
\vspace{-.44cm}
\!\!r_k\!=\!\sum_{m=1}^{M}u_{mk}\mathcal{Q}\!\left(\!\hat{\textbf{g}}_{mk}^{H}\textbf{y}_m\!\right)\!=\!\sum_{m=1}^{M}u_{mk}\!\left(a\hat{\textbf{g}}_{mk}^{H}\textbf{y}_m+{n}_{d,mk} \!\!\right).
 \label{rkcase3}
\end{IEEEeqnarray}
We define $\textbf{u}_k = [u_{1k}, u_{2k},\cdots, u_{Mk}]^T$ without loss of generality, it is assumed that $|| \textbf{u}_k||=1$.
\begin{table}[!t]
\centering 
\caption{The distortion power of a uniform quantizer with Bussgang decomposition, \cite{ourvtc18,ourjournal2}.} 
\vspace{-.31cm}
\label{tabledelta} \label{tablezillmann}
 \begin{tabular}{ c c c c}
 \hline
$\alpha$  & $\Delta_{\text{opt}}$ & $p_{\tilde{n}_d}=\tilde{b}-\tilde{a}^2=\sigma_{\tilde{e}}^2$ & $\tilde{a}$   \\ [.1ex] 
 \hline\hline
 \vspace{.02cm}
{1} & 1.596 &  0.2313 &  0.6366\\[.02ex] 
 \hline
 \vspace{.01cm}
{2} &0.9957 &  0.10472 & 0.88115\\ [.02ex] 
 \hline
  \vspace{.01cm}
{3} &0.586 &  0.036037  &0.96256\\ [.02ex] 
 \hline
  \vspace{.01cm}
{4} &0.3352 & 0.011409  & 0.98845\\ [.02ex] 
 \hline
  \vspace{.01cm}
{5} &0.1881  & 0.003482  & 0.996505\\ [.02ex] 
 \hline	
   \vspace{.01cm}
{6} &0.1041  & 0.0010389  & 0.99896\\ [.02ex] 
 \hline	
   \vspace{.01cm}
{7} &0.0568 & 0.0003042  & 0.99969\\ [.02ex] 
 \hline	
\\
 \end{tabular}
 \vspace{-1cm}
\end{table}
\vspace{-.1cm}
\section{Performance Analysis}
\vspace{-.12cm}
The aggregate received signal at the CPU can be written as

\begin{small}
\begin{IEEEeqnarray}{rCl}
&&\!\!r_k=\tilde{a}\underbrace{\sqrt{\rho} \mathbb{E}\left\{\sum_{m=1}^Mu_{mk}\hat{\textbf{g}}_{mk}^H{\textbf{g}}_{mk}\sqrt{q_k}\right\}}_{\text{DS}_k}s_k\\
 &\!\!+\!\!&
 \tilde{a}\underbrace{\!\sqrt{\rho}\!\left(\!\!\sum_{m=1}^M\!\!u_{mk}\hat{\textbf{g}}_{mk}^H{\textbf{g}}_{mk}\sqrt{q_k}
 \!\!-\!\!
 \mathbb{E}\!\left\{\!\!\sum_{m=1}^M\!\! \!u_{mk}\hat{\textbf{g}}_{mk}^H{\textbf{g}}_{mk}\sqrt{q_k}\!\right\}\!\!\!\right)\!}_{{\text{BU}_k}\!}\!s_k\!\!+\!\!\tilde{a}\!\!\sum_{k^{\prime}\neq k}^{K}\!\!\nonumber\\
 &&\underbrace{\!\sqrt{\rho}\!\!\sum_{m=1}^M\!u_{mk}\hat{\textbf{g}}_{mk}^H{\textbf{g}}_{mk^\prime}\sqrt{q_{k^\prime}}}_{{\text{IUI}_{kk^\prime}}}s_{k^{\prime}}\!\!+\!\!\tilde{a}\underbrace{\!\!\sum_{m=1}^{M}\!u_{mk}\hat{\textbf{g}}_{mk}^H\textbf{n}_m}_{\text{TN}_k} \!\!+\! \!\underbrace{\!\sum_{m=1}^{M}\!u_{mk}n_{d,mk}}_{\text{TQE}_k},\nonumber
 \label{rk2}
\end{IEEEeqnarray}
\end{small}
where $\text{DS}_k$ and $\text{BU}_k$ denote the desired signal (DS) and beamforming uncertainty (BU) for the $k$th user, respectively, and $\text{IUI}_k$ represents the inter-user-interference (IUI) caused by the $k^\prime$th user. In addition, $\text{TN}_k$ accounts for the total noise (TN), and finally $\text{TQE}_k$ refers to the total quantization error (TQE) at the $k$th user.
The elements of quantization error are i.i.d. random variables \cite{Oppenheimsignal}. We exploit a symmetrical quantizer, where the quantization noise has zero mean, if the probability density function of the input of the quantizer is even \cite{max_quantiztion}. 
\begin{proposition}
Using Bussgang decomposition the elements of the quantization error are uncorrelated with the input of the quantizer \cite{Zillmann}, i.e.,
$
\mathbb{E}\left\{\left(\hat{\textbf{g}}_{mk}^{H}\textbf{y}_m\right)^H n_{d,mk}\right\}=0.
$
\end{proposition}
\vspace{-.32cm}
Hence, exploiting the analysis in \cite{marzetta_free16}, it can be shown that terms $\text{DS}_k.s_k$, $\text{BU}_k.s_k$, $\text{IUI}_{kk^\prime}.s_k^\prime$, $\text{TN}_k$ and $\text{TQE}_k$ are mutually uncorrelated.
Using the worst-case Gaussian noise, and the analysis in \cite{marzetta_free16}, the corresponding signal-to-interference-plus-noise ratio (SINR) is
\vspace{-.13cm}
\begin{IEEEeqnarray}{rCl}
\small
\label{sinrdef11}
\!\!\text{SINR}_{k}\!\!=\!\!\!\dfrac{|\text{DS}_k|^2}{\!\!\!\!\!\!\mathbb{E}\!\left\{|\text{BU}_k|^2\right\}\!\!+\!\!\!\!\sum\limits_{k^\prime\ne k}^K\!\!\!\!\!\mathbb{E}\{|\text{IUI}_{kk^\prime}|^2\!\}\!\!+\!\!\mathbb{E}\{|\text{TN}_k|^2\!\}\!\!+\!\!\!\dfrac{1}{\tilde{a}^2}\mathbb{E}\{\!|\text{TQE}_k|^2\!\}\!\!}\!.
\end{IEEEeqnarray}
\begin{figure*}[h]
\begin{IEEEeqnarray}{rCl}
\begin{split}
\!\!\!\!S_k\!\left(q_k,\textbf{u}_k,b\right)\!\approx \!\left(\!\!1-\frac{\tau_p}{\tau_c}\!\right)
\!\!\log_2\!\!\Bigg(\!\! 1\!+\!\dfrac{N^2\textbf{u}_k^H\left(q_k\bgama_k\bgama_k^H\right)\textbf{u}_k}{\textbf{u}_k^H\left(N^2\sum_{k^\prime\ne k}^Kq_{k^\prime}|\pmb{\phi}_k^H\pmb{\phi}_{k^\prime}|^2\bdel_{k k^\prime}\bdel_{k k^\prime}^H+N\sum_{k^\prime=1}^{K}q_{k^\prime}\textbf{D}_{k k^\prime}+\dfrac{N}{\rho}\textbf{R}_{k}\!\!\right )\textbf{u}_k }\Bigg)\!, (\text{bit}/\text{s}/\text{Hz}) 
\label{sinr1}
\end{split}
\end{IEEEeqnarray}
\vspace{-0.94cm}
\end{figure*}
\begin{theorem}
\label{theorem_up_quan_u}
The spectral efficiency of the \textit{k}th user is given by (\ref{sinr1}) (defined at the top of this page), where $\tau_c$ denotes the number of samples for each coherence interval, and
\begin{small}
\begin{subequations}\vspace{-.1cm}
\begin{eqnarray}
&\!\!\!\!\!\!\!\!\!\!\!\!\!\!\!\!\!\!\!\!\!\!\!\!\!\!\!\!\!\!\!\!\!\!\!\!\!\!\!\!\!\!\!\!\!\!\!\!\!\!\!\!\!\!\!\!\!\!\!
\!\!\!\!\!\!\!\!\!\!\!\!\!\!\!\!\!\!
\bgama_k=[\gamma_{1k}, \gamma_{2k}, \cdots, \gamma_{Mk}]^T,\label{eebb1}\\
\vspace{-.3cm}
&
\!\!\!\!\!\!\!\!\!\!\!\!\!\!\!\!\!\!\!\!\!\!\!\!\!\!\!\!\!\textbf{D}_{k k^\prime}\!=\text{diag}\Big[\!\beta_{1k^\prime}\left(\dfrac{\sigma_{\tilde{e}}^2}{\tilde{a}^2}\left(2\beta_{1k}\!-\!\gamma_{1k}\!\right)+\gamma_{1k}\right),\!\cdots\!,\!\label{eebb2}\\
\vspace{-.3cm}
&
~~~~~~~~~~~~\beta_{Mk^\prime}\left(\dfrac{\sigma_{\tilde{e}}^2}{\tilde{a}^2}\left(2\beta_{Mk}-\gamma_{Mk}\!\right)+\gamma_{Mk}\right)\!\Big] \nonumber,
\\
\vspace{-.3cm}
&
\!\!\!\!\!\!\!\!\!\!\!\!\!\!\!\!\!\!\!\!\!\!\!\!\!\!\!\!\!\!\! \bdel_{k k^\prime}=[\dfrac{\gamma_{1k}\beta_{1k^\prime}}{\beta_{1k}}, \dfrac{\gamma_{2k}\beta_{2k^\prime}}{\beta_{2k}}, \cdots, \dfrac{\gamma_{Mk}\beta_{Mk^\prime}}{\beta_{Mk}}]^T\label{eebb3},\\
\vspace{-.3cm}
&
\!\!\!\!\!\!\!\!\!\!\!\!\!\!\!\!\!\!
 \textbf{R}_{k} = \text{diag}\left[\left(\dfrac{\sigma_{\tilde{e}}^2}{\tilde{a}^2}+1\right)\gamma_{1k}, \cdots, \left(\dfrac{\sigma_{\tilde{e}}^2}{\tilde{a}^2}+1\right)\gamma_{Mk}\right]\!\!,\label{eebb4}
 \vspace{-2cm}
\end{eqnarray}
\label{eebb}
\end{subequations}
\end{small}where $\gamma_{mk}=\sqrt{\tau p_p}\beta_{mk}c_{mk}$, and $\text{diag}[\textbf{x}]$ refers to a diagonal matrix whose diagnoal elements are the elements of vector $\textbf{x}$.
\end{theorem}
\vspace{-.2cm}
{\textit{Proof:}} Please refer to Appendix A.~~~~~~~~~~~~~~~~~~~~~~~~~~~~~~~~~~$\blacksquare$
\vspace{-.3cm}
\section{Total Energy Efficiency Model}
 \vspace{-.2cm}
The total power consumption can be defined as follows \cite{emil_ee_updl}: 
\begin{equation}
P_{\text{total}}=P_\text{TX}+P_\text{CP},
\end{equation}
where $P_\text{TX}$ is the uplink power amplifiers (PAs), and  $P_\text{CP}$ refers to the circuit power (CP) consumption. The power consumption $P_\text{TX}$  is given by
$
P_\text{TX}=\frac{1}{\zeta}\rho N_0\sum_{k=1}^Kq_k,
$
where $\zeta$ is the PA efficiency at each user. The power consumption $P_\text{CP}$ is obtained as
$
P_\text{CP} = M P_\text{fix}+K P_\text{U}+\sum_{m=1}^MP_{\text{bh},m},
$
where $P_\text{fix}$ is a fixed power consumption at each AP, $P_{U}$ denotes the required power to run circuit components at each user, and backhaul power consumption from the $m$th AP to the CPU is obtained as follows \cite{fettwis_globe11,ee_backhaul_Imran_trans,ee_backhaul_towards_wcnc}:
\begin{equation}
P_{\text{bh},m}=P_\text{BT}~~\frac{R_{\text{bh},m}}{C_{\text{bh},m}},
\end{equation}
where $P_\text{BT}$ is the total power required for backhaul traffic (BT) at full capacity, $C_{\text{bh},m}$ is the capacity of the backhaul link between the $m$th AP and the CPU, $R_{\text{bh},m}$ is the actual backhaul rate between the $m$th AP and the CPU given by
\begin{equation}R_{\text{bh},m}=\dfrac{2~K~\tau_f~\alpha_m}{T_c},
\end{equation} 
where $\alpha_m$ denotes the number of quantization bits at the $m$th APs. Note that considering the same number of bits at all APs, we drop the index $m$ and use $\alpha$ as the number of quantization bits. Moreover, $\tau_f$ introduces the length of frame (which represents the length of the uplink data, in symbols) and is given by $
\tau_f = \tau_c - \tau_p
$. Hence, the total energy efficiency is given by
\begin{equation}
E_e \left(q_k,\textbf{u}_k,\alpha\right) = \frac{B~S \left(q_k,\textbf{u}_k,\alpha\right)}{P_\text{total}}~(\text{bit}/\text{Joule}),
\end{equation}
where $B$ is the bandwidth.
\vspace{-.14cm}
\section{Energy Efficiency Maximization Scheme}
The total energy efficiency maximization is modelled by
\begin{subequations}
\label{pee1} 
\begin{align}P_1:~~
\label{pee1_1}&\max_{q_k, \textbf{u}_k,\alpha}~   E_e\left(q_k,\textbf{u}_k,\alpha\right) ,\\
\label{pee1_2}\!\!\!\!\!\!\!&\!\!\!\!\!\!\!\text{s.t.} ~S_k\left(q_k,\textbf{u}_k,\alpha\right)\ge S^{(\text{r})}_k, \forall k,0 \le q_k \le p_{\text{max}}^{(k)},  \forall k,\\
\label{pee1_5}&~R_{bh,m} \le C_{\text{bh}},  ~~ \forall m,
\end{align}
\end{subequations}
\vspace{-.03cm}where $S^{(\text{r})}_k$ is the required spectral efficiency of the $k$th user, $p_{\text{max}}^{(k)}$ and $C_{\text{bh},m}$ refer to the maximum transmit power available at user \textit{k} and the capacity of backhaul link between the $m$th AP and the CPU, respectively. Assuming the same amount of backhaul capacity between all APs and the CPU, we drop the index $m$, and use $C_{\text{bh}}$ for simplicity.
Problem $P_1$ contains one discrete variable (the number of quantization bits). Hence, we can formulate the problem for fixed values of the number of quantization bits $\alpha$, and we investigate the optimal values of $\alpha$, in numerical results. As a result, for a given $\alpha$, the total energy efficiency maximization problem can be re-formulated as follows:
\vspace{-.1cm}
\begin{subequations}
\label{pee2} 
\vspace{-.2cm}
\begin{align}P_2:
\label{pee2_1}\!\max_{q_k, \textbf{u}_k}&\!\!\!\!\!\quad \frac{B~.~S \left(q_k,\textbf{u}_k,\alpha\right)}{\frac{1}{\zeta}\rho_dN_0\sum_{k=1}^Kq_k\!+\!M P_\text{fix}\!+\!K P_\text{U}\!+\!P_\text{BT}\frac{2~K~\tau_f~\alpha}{T_c}\frac{P_\text{BT}}{C_\text{bh}}\!} ,\\
\label{pee2_2}\text{s.t. ~}\quad &~~~S_k\left(q_k,\textbf{u}_k,\alpha\right)\ge S^{(\text{r})}_k, ~~ \forall k,~~~~\\
\label{pee2_4}&~~~0 \le q_k \le p_{\text{max}}^{(k)},  ~~ \forall k.
\end{align}
\end{subequations}
We reformulate Problem $P_2$ into the following:
\begin{subequations}
\vspace{-.2cm}
\label{pee3} 
\begin{align}P_3:\\
\vspace{.1cm}
 \vspace{-1.4cm}
\!\max_{q_k, \textbf{u}_k,\nu}&\!\!\!\quad \frac{B~.~S \left(q_k,\textbf{u}_k,\alpha\right)}{\!\!\frac{1}{\zeta}\rho_dN_0\nu\sum_{k=1}^Kp_{\text{max}}^{(k)}\!\!+\!\!M P_\text{fix}\!\!+\!\!K P_\text{U}\!\!+\!\!P_\text{BT}\frac{2~K~\tau_f~\alpha}{T_c}\frac{P_\text{BT}}{C_\text{bh}}\!},\nonumber\\
\label{pee3_2}~~\text{s.t. }\quad &S_k\left(q_k,\textbf{u}_k,\alpha\right)\ge S^{(\text{r})}_k, ~~ \forall k,\\
\label{pee3_4}&\!\!\!\!\!\!\!\!\!\!\!\!\!\!\!\!\!\!\!
0 \le q_k \le p_{\text{max}}^{(k)},\forall k,~\sum_{k=1}^K q_k \le \nu\sum_{k=1}^K p_{\text{max}}^{(k)}, ~\nu^\star \le \nu \le 1,\!\!\!
\end{align}
\end{subequations}
 \vspace{-.1cm}where $\nu$ is a auxiliary variable.
Moreover, based on the analysis in \cite{otterson_ee_15,coordinate_pmp}, the slack variable $\nu^\star$ is obtained by solving a power minimization problem subject to the same per-user power constraints in (\ref{pee2_4}) and throughput requirements in (\ref{pee2_2}). Assuming a total transmit power as $\sum_{k=1}^K q_k$, based on \cite{otterson_ee_15,coordinate_pmp}, the power minimization problem (PMP) can be defined as follows:
\vspace{-.2cm}
\begin{subequations}
\label{pee13} 
\begin{align}
\label{pee13_1}
&\!\!\!\!\!\!\!\!\!P_{\text{PMP}}:\min_{q_k}~~\sum_{k=1}^K q_k\\
\label{pee13_2}&\!\!\!\!\!\!\!\!\!\text{s.t.}~S_k\left(q_k,\textbf{u}_k,\alpha\right)\ge S^{(\text{r})}_k, ~~ \forall k,~\&~ 0 \le q_k \le p_{\text{max}}^{(k)},  \forall k.\!\!\!\!\!\!\!\!
\end{align}
\end{subequations}
\vspace{-.1cm}
Problem $P_{\text{PMP}}$ is a GP and can be efficiently solved. After solving Problem $P_{\text{PMP}}$ and finding the optimal solution $q^{+}_k, \forall k$, the slack variable $\nu^\star$ is obtained as 
$
\nu^\star=\frac{\sum_{k=1}^Kp_{\text{max}}^{(k)}}{\sum_{k=1}^K q^+_k }.
$
$\!\!\!\!$\begin{theorem}\label{theorem_ee_sumrate}
$\!\!$Optimal solution of Problems $\!P_2$ and $\!P_3$ are equal. 
\end{theorem}
\textit{Proof:} Let us assume $\{\textbf{U}^\text{opt},\textbf{q}^\text{opt}\}$ and $\{\tilde{\textbf{U}}^\text{opt},\tilde{\textbf{q}}^\text{opt}, \tilde{\nu}\}$ are the optimal solution of Problems $P_2$ and $P_3$, respectively. It is easy to show that $\sum_{k=1}^K \tilde{q}_k = \tilde{\nu}\sum_{k=1}^K p_{\text{max}}^{(k)}$. Moreover, based on \cite[Theorem 1]{otterson_ee_15}, $\tilde{\textbf{U}}^\text{opt}$ and $\tilde{\textbf{q}}^\text{opt}$ provide a feasible solution to Problem $P_2$. Exploiting the per-user power constraints, using $\nu=\frac{1}{\sum_{k=1}^K p_{\text{max}}^{(k)}}\sum_{k=1}^K q_k$ and $0 \le \nu \le 1$, and by considering the throughput requirements, one can conclude that $\{\textbf{U}^\text{opt},\textbf{q}^\text{opt}\}$ provide a feasible solution to Problem $P_3$. ~~~$\blacksquare$\\\\
Hence, we can convert the original total energy efficiency maximization problem into a throughput maximization problem with the new total power constraint. Next, Problem $P_3$ is iteratively solved by performing a one-dimensional search over the variable $\nu^\star \le \nu \le 1$ \cite{otterson_ee_15}. 
Note that for a given $\nu$, the denominator of the objective function of Problem $P_3$ is a constant. Therefore, we reformulate Problem $P_3$ as follows:
\vspace{-.23cm}
\begin{subequations}
\label{pee4} 
\begin{align}P_4: 
\label{pee4_1}&\max_{q_k, \textbf{u}_k}\quad S \left(q_k,\textbf{u}_k,\alpha\right) ,\\
\label{pee4_2}\!\!\!\!\!\!\!&\!\!\!\!\!\!\!\text{s.t.}\quad ~S_k\left(q_k,\textbf{u}_k,\alpha\right)\ge S^{(\text{r})}_k, \forall k,0 \le q_k \le p_{\text{max}}^{(k)},  \forall k,\\
\label{pee4_5}&~~~~~~~~~\sum_{k=1}^K q_k \le \nu\sum_{k=1}^K p_{\text{max}}^{(k)}.
\end{align}
\end{subequations}
Problem $P_4$ is not jointly convex in terms of $\mathbf{u}_{k}$ and power allocation $q_{k},~\forall k$. To tackle this non-convexity issue, we decouple Problem $P_5$ into two sub-problems: receiver filter coefficient design (i.e. $\mathbf{u}_{k}$) and the power allocation problem. The optimal solution for Problem $P_4$, is obtained through alternately solving these sub-problems, as explained in the following subsections.
\vspace{-.1cm}
\subsection{Receiver Filter Coefficient Design}
We solve the total energy efficiency maximization problem for a given set of power allocations at all users, $q_k, \forall k$, and fixed $\alpha$. These coefficients (i.e., $\mathbf{u}_{k}$, $\forall~k$) are obtained by independently maximizing the total uplink energy efficiency of the system. Hence the optimal receiver filter coefficients are determined by solving the following optimization problem:
\begin{subequations}
\label{pee5} 
\begin{align}P_5:
\label{pee5_1}\!&\max_{\textbf{u}_k}~~~ S_k \left(q_k,\textbf{u}_k,\alpha\right) ,\\
\label{pee5_2}&~\text{s.t. }~~~~S_k\left(q_k,\textbf{u}_k,\alpha\right)\ge S^{(\text{r})}_k, ~~ \forall k.~~~~
\end{align}
\end{subequations}
Note that the satisfaction of constraints in (\ref{pee5_2}) will be ensured in the power allocation problem. Hence, we drop constraint (\ref{pee5_2}) and Problem $P_5$ can be reformulated as:

\begin{small}
\begin{IEEEeqnarray}{rCl}
\label{p6} 
&&P_6:\max_{\mathbf{u}_k}\\
&&\dfrac{\!N^2\textbf{u}_k^H\left(q_k\bgama_k\bgama_k^H\right)\textbf{u}_k\!}{\!\!\textbf{u}_k^H\!\!\left(\!N^2\!\sum_{k^\prime\ne k}^K\!q_{k^\prime}\!|\pmb{\phi}_k^H\pmb{\phi}_{k^\prime}|^2\!\bdel_{k k^\prime}\!\bdel_{k k^\prime}^H\!+\!N\!\sum_{k^\prime=1}^{K}q_{k^\prime}\textbf{D}_{kk^\prime}\!+\!\dfrac{N}{\rho}\!\textbf{R}_{k}\!\right)\!\textbf{u}_k\!}.\nonumber
\end{IEEEeqnarray}
\end{small}
Problem $P_6$ is a generalized eigenvalue problem \cite{bookematrix}, where the optimal solutions can be obtained by determining the generalized eigen vector of the matrix pair $\mathbf{A}_{k} = N^2q_k\bgama_k\bgama_k^H$ and $\mathbf{B}_{k}\!=N^2\sum_{k^\prime\ne k}^K\!q_{k^\prime}\!|\pmb{\phi}_k^H\pmb{\phi}_{k^\prime}|^2\!\bdel_{k k^\prime}\!\bdel_{k k^\prime}^H\!+\!N\sum_{k^\prime=1}^{K}q_{k^\prime}\textbf{D}_{kk^\prime}\!+\!\frac{N}{\rho}\!\textbf{R}_{k}$ corresponding to the maximum generalized eigenvalue.
\vspace{-.1cm}
\subsection{ Power Allocation} 
In this subsection, we solve the power allocation problem for a given set of fixed receiver filter coefficients, $\mathbf{u}_{k}$, $\forall~k$, and fixed values of quantization levels, $Q_m,~\forall m$. The optimal transmit power can be determined by solving the following total energy efficiency maximization problem:
\begin{subequations}
\label{pee7} 
\vspace{-.2cm}
\begin{align}P_7:
\label{pee7_1}&\max_{q_k} ~~~~S \left(q_k,\textbf{u}_k,\alpha\right),\\
\label{pee7_2}\!\!\!\!\!\!\!&\!\!\!\!\!\!\!\text{s.t.}\quad ~S_k\left(q_k,\textbf{u}_k,\alpha\right)\ge S^{(\text{r})}_k, \forall k,0 \le q_k \le p_{\text{max}}^{(k)},  \forall k,\\
\label{pee7_4}&~~~~~~~~~~\sum_{k=1}^K q_k \le \nu\sum_{k=1}^K p_{\text{max}}^{(k)}.
\end{align}
\end{subequations}
As $\log$ is a monotonically increasing function, Problem $P_7$ is reformulated as follows:
\begin{subequations}
\vspace{-.26cm}
\label{pee8} 
\begin{align}
\vspace{-.2cm}
\label{pee8_1}
P_{8}:&\min_{q_k,t_k}~~~~~\prod_{k=1}^K\left(1+t_k\right)^{-1}&\\
\label{pee8_2}\!\!\!\!\!\!\!&\!\!\!\!\!\!\!\text{s.t.}\quad ~S_k\left(q_k,\textbf{u}_k,\alpha\right)\ge S^{(\text{r})}_k, \forall k,0 \le q_k \le p_{\text{max}}^{(k)},  \forall k,\\
\label{pee8_4}&~~~~~~~~~~~\text{SINR}_k \ge t_k,  \forall k,~\sum_{k=1}^K q_k \le \nu\sum_{k=1}^K p_{\text{max}}^{(k)}.
\end{align}
\end{subequations}where $t_k, \forall k$ refers to the slack variables. Problem (\ref{pee8}) is a non-convex signomial problem. However, In Appendix B, we will show that all constraints in (\ref{pee8}) can be reformulated into posynomial functions. Hence, if the objective function in (\ref{pee8_1}) is reformulated into a posynomial function, problem (\ref{pee8}) is a standard GP. This motivates us to propose the following approach to transform Problem (\ref{pee8}) into a standard GP.
We use the SCA scheme proposed in \cite{refrence52ofotterson_ee_15} to approximate Problem (\ref{pee8}) into a standard GP. Based on the analysis in \cite{refrence52ofotterson_ee_15}, it is possible to search for a local optimum through solving a sequence of GPs which locally approximate the original optimization problem. This scheme is called the ``inner approximation algorithm for non-convex problems'' in \cite{refrence52ofotterson_ee_15}. This scheme provides an efficient solution for the original problem \cite{otterson_ee_15,refrence52ofotterson_ee_15}. Next, the following lemma using SCA is required \cite[Lemma 1]{otterson_ee_15}:
\begin{lemma}
\label{lemmalocalappx}
Function $\Theta(x)=\kappa t^\xi$ can be used to approximate function $\Pi(x)=1+t$, near the point $\hat{t}$. The best monomial local approximation is obtained by the following parameters: 
$
\xi=\frac{\hat{t}}{1+\hat{t}},~\kappa=\frac{1+\hat{t}}{\hat{t}^\xi},
$
where $\Theta(t)\le\Pi(t)$, $\forall t>0$.
\end{lemma} 
Using the local approximation in Lemma \ref{lemmalocalappx}, we can tackle the non-convexity of Problem $P_{8}$, which enables us to reformulate Problem $P_{8}$ as follows:
\begin{subequations}
\vspace{-.2cm}
\label{pee9} 
\begin{align}
\vspace{-.2cm}
\label{pee9_1}
P_{9}:&\min_{q_k,t_k}~~~\prod_{k=1}^Kt_k^{-\dfrac{\hat{t}_k}{1+\hat{t}_k}}\\
\label{pee9_2}\!\!\!\!\!\!\!\!\!\!\!\!\!\!\text{s.t.} ~&S_k\left(q_k,\textbf{u}_k,\alpha\right)\ge S^{(\text{r})}_k, \forall k,~~0 \le q_k \le p_{\text{max}}^{(k)},  \forall k,\\
\label{pee9_4}&\text{SINR}_k \ge t_k,  \forall k,~~ \sum_{k=1}^K q_k \le \nu\sum_{k=1}^K p_{\text{max}}^{(k)},\\
\label{pee9_5}&\left((1-\delta)\hat{t}_k\right) \le t_k \le \left((1-\delta)\hat{t}_k\right),  \forall k,
\end{align}
\end{subequations}
where $\delta=0.1$ and it controls the approximation accuracy \cite{otterson_ee_15}.
\begin{proposition}\label{prop_ee_opt}
Problem $P_{9}$ is formulated into a standard GP. 
\end{proposition}
{\textit{Proof:}} Please refer to Appendix B.~~~~~~~~~~~~~~~~~~~~~~~~~~~~~~~~~$\blacksquare$
\\\\
Therefore, Problem $P_{9}$ is efficiently solved through existing convex optimization software. Based on these two sub-problems ($P_6$ and $P_{9}$), iterative  Algorithm \ref{eeal1} has been developed by alternately solving both sub-problems, where we set $\epsilon_1=\epsilon_2=0.01$.
\begin{algorithm}[h]
\caption{Proposed algorithm to solve Problem $P_4$}\vspace{-.07cm}
\label{eeal1}
\hrulefill

\textbf{1.} Initialize $\textbf{q}^{(0)}$, $\textbf{U}^{(0)}$. Calculate the uplink $\text{SINR}_k^{(0)}$, $t_0^{(0)}$ and $S_k^{(r)}$ using $\textbf{q}^{(0)}$ and $\textbf{U}^{(0)}$, and set the initial SINR guess and initial auxiliary variables as $\hat{t}_k=\text{SINR}_k^{(0)}, \forall k$, and $t_k^{(0)}=\text{SINR}_k^{(0)}, \forall k$, respectively. 

\textbf{2.} Set $\textbf{q}^{(\star)}=0$, $t_k^{(\star)}=t_k^{(0)}$, $\textbf{U}^{(\star)}=\textbf{U}^{(0)}$, and $\tilde{E}_{e,k}^{(\star)}=0,\forall k$.

\textbf{3.} Calculate the constants $\xi$ and $\kappa$ using Lemma \ref{lemmalocalappx}, and solve Problem $P_{9}$ with $t_k^{(\star)}$ and $\textbf{U}^{(\star)}$, and find $\textbf{q}^{(\star\star)}$ and calculate $t_0^{(\star\star)}$ and $t_k^{(\star\star)}$.
 
\textbf{4.} If $\left| t_k^{(\star\star)}-t_k^{(\star)}\right|\le 0.01$, then set $t_k^{(\star\star)}=t_k^{(\star)}$ and $\textbf{q}^{(\star\star)}=\textbf{q}^{(\star)}$ and go to step 8, otherwise, $t_k^{(\star)}=t_k^{(\star\star)}$ and go to step 3.

\textbf{5.} Solve Problem $P_6$ using $\textbf{q}^{(\star)}$ and calculate $\textbf{U}$ and set $\textbf{U}^{(\star\star)}=\textbf{U}$.

\textbf{6.} Compute the objective value of Problem $P_{9}$ with $\textbf{U}^{(\star\star)}$ and $\textbf{q}^{(\star)}$ and call it $\tilde{E}_{e,k}^{(\star\star)},\forall k$.

\textbf{7.} If $\left| \tilde{E}_{e,k}^{(\star\star)}-\tilde{E}_{,k}e^{(\star)}\right|\le 0.01,\forall k$, then $\textbf{U}^{(\star)}=\textbf{U}^{(\star\star)}$ and go to step 8, otherwise, go to step 3.

{\textbf{8.$\!$} If the stopping criteria is satisfied terminate, otherwise, go to step $\!$3. }
\vspace{-.2cm}
\end{algorithm}
\section{Numerical Results}
A cell-free Massive MIMO system with $M$ APs and $K$ single-antenna users is considered in a $D \times D$ simulation area, where both APs and users are uniformly distributed
at random. An uncorrelated shadowing model and a three-slope model for the path loss similar to \cite{marzetta_free16} are considered. It is assumed that that $\bar{p}_p$ and $\bar{\rho}$ denote the power of the pilot sequence and the uplink data powers, respectively, where $p_p=\frac{\bar{p}_p}{p_n}$ and $\rho=\frac{\bar{\rho}}{p_n}$, and we set $\bar{p}_p=200$ mW and $\bar{\rho}=1$ Watt, unless otherwise indicated. In addition, $p_n$ refers to the noise power and we exploited the analysis in \cite{marzetta_free16} to calculate it.
Moreover, we use $\zeta=0.3$, $P_U=0.1$ Watt, $P_{\text{fix}}=0.825$ Watt \cite{fettwis_globe11,emil_ee_updl,ee_backhaul_Imran_trans,ee_backhaul_towards_wcnc}.
\begin{figure}[t!]
\center
\includegraphics[width=73mm]{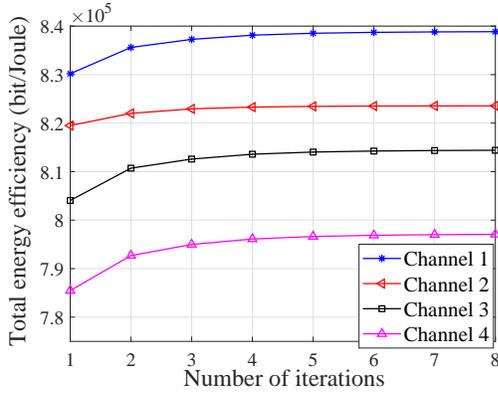}
\vspace{-.36cm}
\caption{The total energy efficiency of proposed Algorithm 1 versus number of iterations with $K=20$, $M=100$, $N=1$, $\alpha=2$, $\tau_p=20$, and $D=1$ km.}
\label{conv_k20_m100_n1_b2_rho1_t20_local_glob}
\end{figure}
\vspace{-.1cm}
\begin{figure}[t!]
\center
\includegraphics[width=73mm]{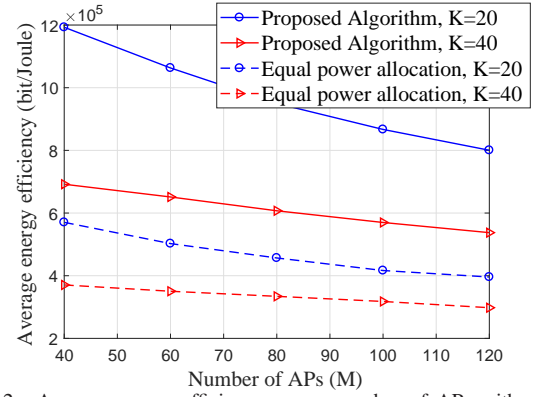} 
\vspace{-.35cm}
\caption{Average energy efficiency versus number of APs with proposed Algorithm 1 and equal power allocation with $M=100$, $N=1$, $\alpha=2$, $\tau_p=20$, and $D=1$ km.}
\label{EE_Prod_wo_vsM_K2040_t20_glob}
\end{figure}
\begin{figure}[t!]
\center
\includegraphics[width=73mm]{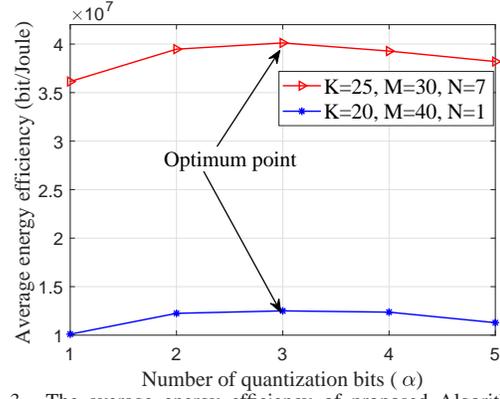}
\vspace{-.35cm}
\caption{The average energy efficiency of proposed Algorithm 1 versus number of quantization bits with $K=20$, $N=1$, $\tau_p=20$, and $D=1$ km.}
\label{conf_vs_alpha_k2520_m3040_n71_ortho_d1_rho1_pt1}
\end{figure}
\begin{figure}[t!]
\center
\includegraphics[width=73mm]{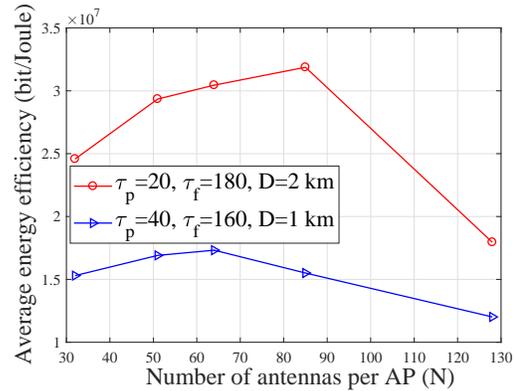}
\vspace{-.39cm}
\caption{The average energy efficiency of proposed Algorithm 1 versus the number of antennas per AP with $MN=256$, $K=40$, $P_{BT}=10$ Watt, $C_{bh}=100$ Mbps, and $\alpha=4$ bits.}
\label{mn_k40_rho1_q4_pt10_d1_d2_t20_t40_glob}
\end{figure}
First, the convergence of the proposed Algorithm \ref{eeal1} is investigated.  
Fig. \ref{conv_k20_m100_n1_b2_rho1_t20_local_glob},  presents the convergence of the proposed algorithm with $M=100$, $K=20$, $N=1$, $\tau_p=20$, and $\alpha=2$. The figure confirms that the proposed Algorithm \ref{eeal1} converges in a few iterations. 
Fig. \ref{EE_Prod_wo_vsM_K2040_t20_glob} presents the total energy efficiency of the proposed Algorithm \ref{eeal1} and the scheme with the equal power allocation with $M=100$, $N=1$, $\alpha=2$, $\tau_p=20$, and $D=1$ km. As seen in Fig. \ref{EE_Prod_wo_vsM_K2040_t20_glob}, the proposed scheme significantly improves the total energy efficiency of cell-free Massive MIMO compared to equal power allocation scheme (i.e., $q_k=1,\forall k, \textbf{u}_k=[1,\cdots,1],\forall k$).
In Fig. \ref{conf_vs_alpha_k2520_m3040_n71_ortho_d1_rho1_pt1}, we investigate the effect of number of quantization bits on the average energy efficiency performance of the system with orthogonal pilots, $P_{\text{BT}}=1$ Watt and $D=1$. By increasing number of quantization bits the spectral efficiency of the system increases, however, at the same time the required capacity of backhaul links increases which results in an optimum point in Fig. \ref{conf_vs_alpha_k2520_m3040_n71_ortho_d1_rho1_pt1} which maximizes the energy efficiency of the system. 
In Fig. \ref{mn_k40_rho1_q4_pt10_d1_d2_t20_t40_glob}, we set $MN=256$ as the total number of service antennas, and it can be seen for a fixed total number of service antennas, by reducing the total number of APs, $M$ (which is equivalent to increasing number of antennas per APs, $N$), the total power consumption will decrease. On the other hand, reducing $M$ results in throughput reduction. As a result, one can find a trade off between $M$ and $N$. Fig. \ref{mn_k40_rho1_q4_pt10_d1_d2_t20_t40_glob} reveals the optimum values of $M$ and $N$ to have the highest total energy efficiency.
\vspace{-.18cm}
\section{Conclusions}
\vspace{-.18cm}
We have considered cell-free Massive MIMO and analysed (using the Bussgang theorem) the scenario when a quantized version of the MRC weighted signals are available at the CPU. Per-user power constraints, backhaul capacity constraints and throughput requirements have been considered and an SCA has been exploited to convert the power allocation problem into a GP and efficiently solve the non-convex problem. Numerical results confirm that the proposed limited-backhaul system, while satisfying the optimization constraints, can achieve almost twice the uplink total energy efficiency compared to the case of equal power allocation. In addition, we examined the trade-off between the total number of APs and the number of antennas at the APs,  for a given total number of antennas, and found that there is an optimal number of AP antennas which depends on the system parameters. Finally, the optimal number of quantization bits to maximize the uplink total energy efficiency has been determined.
\vspace{-.18cm}
\section*{Appendix A: Proof of Theorem \ref{theorem_up_quan_u}} 
The desired signal for the user $k$ is given by
\begin{small}
\begin{IEEEeqnarray}{rCl}
\vspace{-.1cm}
\!\!\!\!\text{DS}_k\!\!=\!\!\sqrt{\rho}\mathbb{E}\!\left\{\!\sum_{m=1}^{M}\!u_{mk}\hat{\textbf{g}}_{mk}^H\textbf{g}_{mk}\!\sqrt{q_k}\right\}\!
=N\sqrt{\rho q_k}\!\sum_{m=1}^{M}\!u_{mk}\!\gamma_{mk}.
\label{dsk_vector}
\end{IEEEeqnarray}
\end{small}
\vspace{-.1in}
The term $\mathbb{E}\{\left | \text{BU}_k\right |^2\}$ can be obtained as
\vspace{-.1cm}
\begin{IEEEeqnarray}{rCl}
\small
&&\mathbb{E} \left\{ \left | \text{BU}_k \right | ^2\right\} = \rho\mathbb{E}  \Biggl
\{ \Biggl| \sum_{m=1}^Mu_{mk}\hat{\textbf{g}}_{mk}^H{\textbf{g}}_{mk}\sqrt{q_k}-\\
\vspace{-.01cm}
 &\!\!\!\!\!&\rho \mathbb{E}\Big\{\sum_{m=1}^Mu_{mk}\hat{\textbf{g}}_{mk}^H{\textbf{g}}_{mk}\sqrt{q_k}\Big\}\Biggr|^2 \Biggr \} \!=\!\rho N\sum_{m=1}^Mq_ku_{mk}^2\gamma_{mk}\beta_{mk},\nonumber
\label{ebuk}
\end{IEEEeqnarray}
\vspace{-.08cm}
where the last equality comes from the analysis in \cite{marzetta_free16}.
The term $\mathbb{E}\{\left | \text{IUI}_{k k^\prime}\right |^2\}$ is obtained as 

\begin{IEEEeqnarray}{rCl}
\small
\mathbb{E}& \{| &\text{IUI}_{k k^\prime} |^2 \}= \rho  \underbrace{q_{k^\prime} \mathbb{E}\left \{\left |\sum_{m=1}^Mc_{mk}u_{mk}\textbf{g}_{mk^\prime}^H\tilde{\bf{w}}_{mk}  \right |^2\right\}}_{A}
\nonumber\\
 &\!\!\!\!\!\!\!\!\!\!\!\!\!\!\!+&\!\!\!\!\!\!\!\rho \underbrace{\!\tau p_p\mathbb{E}\! \left \{ \!q_{k^\prime}\!\left | \! \sum_{m=1}^M\!c_{mk}u_{mk}\!\Big(\sum_{i=1}^{K}\textbf{g}_{mi}\!\pmb{\phi}_k^H
 \pmb{\phi}_i\!\Big)^H{\textbf{g}}_{mk^\prime}\!\right |^2 \right \} }_{B}.
 \label{eiui}
\end{IEEEeqnarray}
\vspace{-.08cm}
Since $\tilde{\bf{w}}_{mk}=\pmb{\phi}_k^H\bf{W}_{p,m}$ is independent from the term $g_{mk^\prime}$ similar to \cite[Appendix A]{marzetta_free16}, the term $A$ in (\ref{eiui}) immediately is given by
$
A = N q_{k^\prime} \sum_{m=1}^Mc_{mk}^2u_{mk}^2\beta_{mk^\prime}.
$
The term $B$ in (\ref{eiui}) can be obtained as
\begin{IEEEeqnarray}{rCl}
 \label{bb111} 
B &=&  \underbrace{\tau p_p q_{k^\prime}\mathbb{E}\left \{\left | \sum_{m=1}^M c_{mk}u_{mk} ||{\textbf{g}}_{mk^\prime}||^2\pmb{\phi}_k^H{\pmb{\phi}}_{k^\prime}\right |^2\right \}}_{C}\\
\vspace{-.1cm}
 &+&\underbrace{\tau p_p q_{k^\prime}\mathbb{E} \left\{\left | \sum_{m=1}^Mc_{mk}u_{mk} \Big(\sum_{i\ne k^\prime}^{K}\textbf{g}_{mi}\pmb{\phi}_k^H\pmb{\phi}_i\Big)^H {\textbf{g}}_{mk^\prime}\right |^2 \right\}}_{D}.\nonumber
 \vspace{-.01cm}
\end{IEEEeqnarray}
\vspace{-.005cm}
The first term in (\ref{bb111}) is given by
\vspace{-.18cm}
\begin{IEEEeqnarray}{rCl}
\vspace{-.2cm}
C &=&N \tau p_p q_{k^\prime}\left |\pmb{\phi}_k^H{\pmb{\phi}}_{k^\prime}\right |^2\sum_{m=1}^Mc_{mk}^2u_{mk}^2\beta_{mk^\prime}\nonumber\\
 &+&
 N^2q_{k^\prime}\left |\pmb{\phi}_k^H{\pmb{\phi}}_{k^\prime}\right |^2\left(\sum_{m=1}^M u_{mk}\gamma_{mk}\dfrac{\beta_{mk^\prime}}{\beta_{mk}}\right)^2,
 \label{CC}
 \vspace{-.2cm}
\end{IEEEeqnarray}
and
\begin{IEEEeqnarray}{rCl}
\small
D 
 &=&\!N\!\sqrt{\tau p_p}q_{k^\prime}\!\sum_{m=1}^{M}\!u_{mk}^2c_{mk}\beta_{mk^\prime}\beta_{mk}\!-\!Nq_{k^\prime}\!\sum_{m=1}^{M}\!u_{mk}^2c_{mk}^2\beta_{mk^\prime}\nonumber\\
 &-&N\tau p_p q_{k^\prime}\sum_{m=1}^{M}u_{mk}^2c_{mk}^2\beta_{mk^\prime}^2\left| \pmb{\phi}_k^H{\pmb{\phi}}_{k^\prime}\right|^2.
 \label{d}
\end{IEEEeqnarray} 
Finally by substituting (\ref{CC}) and (\ref{d}) into (\ref{bb111}), and substituting (\ref{bb111}) into (\ref{eiui}), we obtain
\vspace{-.3cm}
\begin{IEEEeqnarray}{rCl}
 \label{euiu}
\mathbb{E}\{| \text{IUI}_{k k^\prime}|^2\} &=&N\rho q_{k^\prime}\left(\sum_{m=1}^{M}u_{mk}^2\beta_{mk^\prime}\gamma_{mk}\right)\\
 &+&N^2 \rho q_{k^\prime} \left|\pmb{\phi}_k^H{\pmb{\phi}}_{k^\prime}\right|^2 \left(\sum_{m=1}^{M}u_{mk} \gamma_{mk}\dfrac{\beta_{mk^\prime}}{\beta_{mk}}\right)^2.\nonumber
\end{IEEEeqnarray}
The total noise for the user $k$ is given by
\vspace{-.17cm}
\begin{IEEEeqnarray}{rCl}
\!\!\!\!\!\!\!\mathbb{E}\!\left\{\!\left|\!\text{TN}_k\right|^2\!\right\}\!=\!\mathbb{E}\!\left\{\left|\!\sum_{m=1}^{M}u_{mk}\hat{\textbf{g}}_{mk}^H\textbf{n}_m\!\right|^2\!\right\}\!=\!N\sum_{m=1}^{M}u_{mk}^2\gamma_{mk},\!
\label{tn}
\end{IEEEeqnarray}
where the last equality is due to the fact that the terms $\hat{\textbf{g}}_{mk}$ and $\textbf{n}_m$ are uncorrelated.
The power of the quantization error for user $k$ is given by
\vspace{-.21cm}
\begin{IEEEeqnarray}{rCl}
\!\!\!\!\!\!\!\!\!\!\!\mathbb{E}\!\!\left\{\!\!\left|\!\text{TQE}_k\!\right|^2\!\right\}\!\!=\!\!\!\mathbb{E}\!\left\{\!\left|\sum_{m=1}^{M}u_{mk}n_{d,mk}\right|^2\!\!\right\}\!\!=\!\!\!\sum_{m=1}^{M}\!\!u_{mk}^2\!\mathbb{E}\!\left\{\left|n_{d,mk}\right|^2\right\}\!\!,\!
\label{tq}
\end{IEEEeqnarray}
where the last equality is due to the fact that the elements of $e_{mk}$ and $u_{mk}$ are uncorrelated. 
Next, we use the following property of the quantization distortion power 
$
\mathbb{E}\left\{\left|n_{d,mk}\right|^2\!\right\} = \sigma_{\hat{\textbf{g}}_{mk}^{H}\textbf{y}_m}^2\mathbb{E}\left\{\left|{\tilde{n}}_{d,mk}\right|^2\right\}.
$
Defining $z_{mk}=\hat{\textbf{g}}_{mk}^{H}\textbf{y}_m$, we have
\vspace{-.12cm}
\begin{IEEEeqnarray}{rCl}\label{e1}
&&\sigma_{z_{mk}}^2 = \mathbb{E}\left\{ \left(\hat{\textbf{g}}_{mk}^{H}\textbf{y}_m\right)^H\left(\hat{\textbf{g}}_{mk}^{H}\textbf{y}_m\right)\right\}\\
 &\!\!\approx\!&\rho\mathbb{E}\left\{\!\left|\!\sum_{k^{\prime}=1}^{K}u_{mk}\hat{\textbf{g}}_{mk}^H\textbf{g}_{mk^\prime}\sqrt{q_{k^\prime}}s_{k^\prime}\! \right|^2\!\right\}\!\!+\!\mathbb{E}\left\{\left| u_{mk}\hat{\textbf{g}}_{mk}^H\textbf{n}_m \!\right|^2\!\right\}.\nonumber
\end{IEEEeqnarray}For the second term of (\ref{e1}), we have
$
\mathbb{E}\left\{ \left| \hat{\textbf{g}}_{mk}^H\textbf{n}_m \right|^2 \right\} = N \gamma_{mk}.
$
The first term in (\ref{e1}) can be obtained as

\begin{small}
\begin{IEEEeqnarray}{rCl}
&&\!\!\!\mathbb{E}\Bigg\{ \left| \sum_{k^{\prime}=1}^{K}u_{mk}\hat{\textbf{g}}_{mk}^H\textbf{g}_{mk^\prime}\sqrt{q_{k^\prime}}s_{k^\prime} \right|^2 \Bigg\}\\
 \!&\!\!&\!\!\!\!\!\!\!\underbrace{\mathbb{E}\!\Bigg\{\!\!\left|\!\sum_{k^{\prime}=1}^{K}\!u_{mk}{\textbf{g}}_{mk}^H\textbf{g}_{mk^\prime}\sqrt{q_{k^\prime}}s_{k^\prime} \!\right|^2\!\Bigg\}}_{\text{I}}\!\!+\!\!
 \underbrace{\mathbb{E}\!\Bigg\{\!\left|\!\sum_{k^{\prime}=1}^{K}u_{mk}{\bep}_{mk}^H\textbf{g}_{mk^\prime}\sqrt{q_{k^\prime}}s_{k^\prime}\!\right|^2\!\!\Bigg\}\!}_{\text{II}},\nonumber
\label{e3}
\end{IEEEeqnarray}
\end{small}
 where each element of $\bep$ is given by $\epsilon_{mk}= \mathcal{CN}(0,\beta_{mk}-\gamma_{mk})$. The terms I and II in (\ref{e3}) are given as follows:
$
\text{I} = N \beta_{mk}\sum_{k^\prime=1}^{K}q_{k^\prime}\beta_{mk^\prime},
$
and
$
\text{II} = N (\beta_{mk}-\gamma_{mk})\sum_{k^\prime=1}^{K}q_{k^\prime}\beta_{mk^\prime}.
$
Next, let us assume, using the Bussgang decomposition, that we have $\mathbf{v_k}=\mathcal{Q}(\mathbf{z}_k)=a\mathbf{z}_k+\mathbf{d}_k^z$, where $\mathbf{z}_k=\left[z_{1k}\cdots z_{Mk} \right]$ and $z_{mk}=\hat{\textbf{g}}_{mk}^{H}\textbf{y}_m$, where $z_{mk}$ is the input of the quantizer at the $m$th AP. In addition, $\mathbf{d}_k^z=\left[d_{1k}^z\cdots d_{Mk}^z \right]$, and $a$ and $\mathbf{d}_k^z$ are the Bussgang scalar factor and the quantization error, respectively, as defined in Subsection \ref{sec_ee_bussa}. Based on the analysis in \cite{mezghani_WSA16}, we have
\vspace{-.2cm}
\begin{IEEEeqnarray}{rCl}
 \mathbf{R}_{d_k^zd_k^z}
 \stackrel{(a)}{\approx}\left(b-a^2\right)\text{diag}(\mathbf{R}_{z_kz_k}),
 \end{IEEEeqnarray}where $\mathbf{R}_{d_k^zd_k^z}$ and $\mathbf{R}_{z_kz_k}$ refer to the covariance matrix of the quantization error and the covariance matrix of the input of the  quantizer, respectively. Moreover, note that in step (a), we exploit the analysis in  \cite[Section V]{mezghani_WSA16}. Thus, we have:
\begin{IEEEeqnarray}{rCl}
\mathbb{E}\left\{\left|\sum_{m=1}^{M}u_{mk}d_{mk}^z\right|^2\right\}
\approx
\sum_{m=1}^{M}u_{mk}^2\mathbb{E}\left\{\left|d_{mk}^z\right|^2\right\}.
\label{tq}
\end{IEEEeqnarray}
Finally, exploiting (\ref{e1}) and (\ref{tq}), we have
\begin{small}
\begin{IEEEeqnarray}{rCl}
\!\!\!\mathbb{E}\!\left\{\!\left|\!\text{TQE}_k\!\right|^2\!\right\}
\!\!\approx\!\!
N\underbrace{\!\left(\!\tilde{b}\!-\!\tilde{a}^2\!\right)}_{\sigma_{\tilde{e}}^2}\!\!\sum_{m=1}^M\!\! u_{mk}^2 \!\!\left(\!\!\rho\!\left(\!2\beta_{mk}\!-\!\gamma_{mk}\!\right)\!\!\sum_{k^\prime=1}^{K}\!q_{k^\prime}\!\beta_{mk^\prime}\!\!+\! \!\gamma_{mk}\!\!\!\right)\!\!.~~~~
\label{e6}
\end{IEEEeqnarray}
\end{small}
\vspace{-.01cm} 
By substituting (\ref{dsk_vector}), (\ref{ebuk}), (\ref{euiu}) and (\ref{tn}) into (\ref{sinrdef11}), the corresponding spectral efficiency of the $k$th user is obtained by (\ref{sinr1}), which completes the proof of Theorem \ref{theorem_up_quan_u}.
~~~~~~~~~~~~~~~~~~~~~~~~~~~~$\blacksquare$
\vspace{-.24cm}
\section*{Appendix B: Proof of Proposition \ref{prop_ee_opt}}
\vspace{-.15cm} 
The standard form of GP is defined as follows \cite{bookboyd}:
\vspace{-.1cm} 
\begin{subequations}
\label{pee14} 
\begin{align}P_{\text{GP}}:~~~~
\label{pee14_1}&\min   f_0(\textbf{x}),\\
\label{pee14_2}\!\!\!\!\!\!\!\!\!\!\!\!\!\!\!\!\!\!\!\!\!\!\!\!&\!\!\!\!\!\!\!\!\!\!\!\!\!\!\!\!\!\!\!\!\!\!\!\!\!\text{s.t.}~ f_i(\textbf{x}) \le 1,i=1,\cdots,m, ~g_i(\textbf{x}) = 1, ~~i=1,\cdots,p,
\end{align}
\end{subequations}
where $f_0$ and $f_i$ are posynomial and $g_i$ are monomial. Moreover, $\textbf{x}=\{x_1,\cdots,x_n\}$ is the optimization variables. The SINR constraint in (\ref{pee14}) is not a posynomial function, however it can be rewritten into the following posynomial function:

\begin{small}
\begin{IEEEeqnarray}{rCl}
\dfrac{\!\textbf{u}_k^H\!\!\!\left(\!\!N^2\!\sum_{k^\prime\ne k}^Kq_{k^\prime}|\pmb{\phi}_k^H\pmb{\phi}_{k^\prime}|^2\!\bdel_{k k^\prime}\bdel_{k k^\prime}^H\!\!+\!\!N\!\sum_{k^\prime=1}^{K}q_{k^\prime}\textbf{D}_{k k^\prime}\!\!+\!\!\dfrac{N}{\rho}\textbf{R}_{k}\!\!\!\right )\!\!\textbf{u}_k}{\mathbf{u}_k^H\!\left(N^2\!q_k\bgama_k\bgama_k^H\!\right)\!\mathbf{u}_k}&&\nonumber \\ && \!\!\!\!\! \!\!\!\!\!\!\!\!\!\!\!\!\!\!\!\!\!\!\!\!\!\!\!\!\! \!\!\!\!\! \le\!\dfrac{1}{t},~\forall k.~~~~~
\label{ee_inv}
\end{IEEEeqnarray}

\end{small}
By applying a simple transformation, (\ref{ee_inv}) is equivalent to the following inequality:
\vspace{-.2cm}
\begin{IEEEeqnarray}{rCl}
q_k^{-1}\left(\sum_{k^\prime\ne k}^K\!a_{kk^\prime}q_{k^\prime}\!+\!\sum_{k^\prime=1}^{K}b_{kk^\prime}q_{k^\prime}+c_k\right) \le\dfrac{1}{t},
\label{ee_inv2}
\end{IEEEeqnarray}
\\
where 
$a_{kk^\prime}=\frac{\mathbf{u}_k^H (|\!\pmb{\phi}_k^H\!\pmb{\phi}_{k^\prime}|^2\!\bdel_{k k^\prime}\!\bdel_{k k^\prime}^H) \mathbf{u}_k}{\mathbf{u}_k^H\!\left(\bgama_k\bgama_k^H\right)\mathbf{u}_k}$, 
$b_{kk^\prime}=\frac{\mathbf{u}_k^H \textbf{D}_{k k^\prime}\mathbf{u}_k}{\mathbf{u}_k^H(N\bgama_k\bgama_k^H)\mathbf{u}_k}$,
$c_k=\frac{\mathbf{u}_k^H \textbf{R}_{k} \mathbf{u}_k}{\mathbf{u}_k^H(\rho N\bgama_k\bgama_k^H)\mathbf{u}_k}$.
The transformation in (\ref{ee_inv2}) shows that the left-hand side of (\ref{ee_inv}) is a posynomial function. Similarly, it can be shown that the spectral efficiency constraint in (\ref{pee8_2}) can be transformed to a posynomial function.
~~~~~~~~~~~~~~~~~~~~~$\blacksquare$
\bibliographystyle{IEEEtran}
\bibliography{icc_ee_camera_4_cuma} 
\end{document}